\documentclass[twocolumn,preprintnumbers,amsmath,amssymb,superscriptaddress]{revtex4}

\usepackage{graphicx}
\usepackage{dcolumn}
\usepackage{bm}

\begin{document}

\title{Spin Order by Frustration in Triangular Lattice Mott Insulator NaCrO$_2$ : A Neutron Scattering Study}

\author{D.~Hsieh}
\affiliation{Joseph Henry Laboratories of Physics, Princeton University, Princeton, New Jersey 08544, USA}
\affiliation{(Present Address)Institute for Quantum Information and Matter, California Institute of Technology, Pasadena, California 91125, USA}
\author{D.~Qian}
\affiliation{Joseph Henry Laboratories of Physics, Princeton University, Princeton, New Jersey 08544, USA}
\affiliation{(Present Address)Department of Physics and Astronomy, Shanghai Jiao Tong University, Shanghai 200240, China}
\author{R.~F.~Berger}
\affiliation{Department of Chemistry, Princeton University, Princeton, New Jersey 08544, USA}
\author{Chang~Liu}
\affiliation{Joseph Henry Laboratories of Physics, Princeton University, Princeton, New Jersey 08544, USA}
\author{B.~Ueland}
\affiliation{Department of Physics, Pennsylvania State University, University Park, Pennsylvania 16802, USA}
\author{P.~Schiffer}
\affiliation{Department of Physics, Pennsylvania State University, University Park, Pennsylvania 16802, USA}
\author{Q.~Huang}
\affiliation{NIST Center for Neutron Research, Gaithersburg, Maryland 20899, USA}
\author{R.~J.~Cava}
\affiliation{Department of Chemistry, Princeton University, Princeton, New Jersey 08544, USA}
\author{J.~W.~Lynn}
\affiliation{NIST Center for Neutron Research, Gaithersburg, Maryland 20899, USA}
\author{M.~Z.~Hasan}
\affiliation{Joseph Henry Laboratories of Physics, Princeton University, Princeton, New Jersey 08544, USA}

\date{\today}

\begin{abstract}
We report high resolution neutron scattering measurements on the
triangular lattice Mott insulator Na$_x$CrO$_2$ ($x$=1) which has
recently been shown to exhibit an unusually broad fluctuating
crossover regime extending far below the onset of spin freezing
($T_c\sim$41K). Our results show that below some crossover
temperature ($T\sim0.75T_c$) a small incommensuration develops which
helps resolve the spin frustration and drives three-dimensional
magnetic order supporting coherent spin wave modes. This
incommensuration assisted dimensional crossover suggests that
inter-layer frustration is responsible for stabilizing the rare 2D
correlated phase above 0.75$T_c$. In contrast to the host compound
of 2D cobaltate superconductor such as Na$_x$CoO$_2$
($\xi_c>50$\AA), no magnetic long-range order is observed down to
1.5K ($\xi_c<16$\AA).
\end{abstract}


\maketitle

Two dimensional (2D) systems such as rare-gas atoms adsorbed on
graphite \cite{Birgenau}, quantum wells \cite{Klitzing} or graphite
intercalation compounds \cite{Wiesler} are routinely engineered in
the laboratory to search for spontaneously emerging 2D correlated
phases. However spontaneous emergence of a 2D correlated phase in a
bulk 3D system is rare and can arise because of geometrical
frustration. Specific examples are the square lattice
antiferromagnets (AFM) belonging to the rare-earth copper-oxides
\cite{Lynn}, and the spin dimer system BaCuSi$_2$O$_6$
\cite{Sebastian}. Recent discoveries of possible low temperature
spin liquid phases in Cs$_2$CuCl$_4$ \cite{Coldea} and NiGa$_2$S$_4$
\cite{Nakatsuji}, a rich phase diagram for Na$_x$CoO$_2$ \cite{Foo},
and exotic multiferroic behavior in \textit{A}CrO$_2$ systems
\cite{Cheong, Seki} have generated interest in whether geometrical
frustration might stabilize novel 2D phases on stacked triangular
lattices (TL). Such a state has been theoretically proposed to
realize the long sought $Z_2$ topological vortex phase
\cite{Topological}.

Early neutron diffraction work on the TL Mott insulator NaCrO$_2$
demonstrated highly anisotropic magnetic correlations developing
near $T\sim$45K which persist down to 2K \cite{Soubeyroux}. Recent
$\mu$SR and NMR measurements \cite{Olariu} reveal an onset of
gradual spin freezing at $T_{c}\sim$41K, followed by an unusually
broad fluctuation regime that reaches a maximal spin relaxation rate
at 0.75$T_c$. Although strong fluctuations are a hallmark of highly
frustrated systems, the relationship between frustration and the 2D
static behavior is unknown because detailed knowledge of the
magnetic structure is lacking. Here we study the temperature
evolution of both the static and dynamic spin correlations with high
resolution neutron scattering measurements, and find that
inter-layer frustration is essential for stabilizing the rare 2D
antiferromagnetically correlated phase in NaCrO$_2$.

\begin{figure}
\includegraphics[scale=0.3,clip=true, viewport=0.0in 0.0in 9.9in
6.5in]{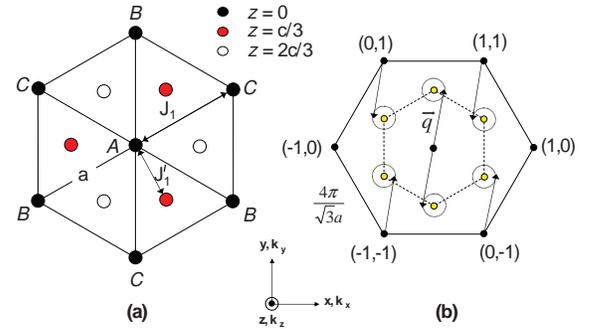} \caption{Crystal structure of NaCrO$_2$.
The \textit{R}$\bar{3}$m crystal symmetry of NaCrO$_2$ was verified
at both 300 K and 4 K by Rietveld refinement of high resolution
neutron powder diffraction data. Lattice constants at 4 K are $a$ =
2.968 \AA\ and $c$ = 15.944 \AA. (a) [001] view of Cr sublattice.
The in-plane $J_{1}$ and inter-plane $J_1'$ exchange paths are
marked, as well as the three-sublattice ($ABC$) structure
corresponding to $\vec{q}=(0,4\pi/3a,0)$. (b) Reciprocal lattice of
hexagonal crystal structure (black dots) and three-sublattice
magnetic structure (yellow dots). Measured low temperature $\vec{q}$
(arrows) deviates slightly from $(0,4\pi/3a,0)$.}
\label{fig:NaCrO2_Fig1'}
\end{figure}

The classical ground state of the nearest neighbor TLAFM is the
120$^{\circ}$ spin structure. If such layers are rhombohedrally
stacked [fig.~\ref{fig:NaCrO2_Fig1'}(a)], there is a cancellation of
inter-layer Weiss fields which leads to a classical decoupling of
spin layers. The $A$CrO$_2$ ($A$=Li, Na, K) partially-filled band
Mott insulators, in which Cr$^{3+}$ ($S=\frac{3}{2}$) ions form a
rhombohedrally stacked TL [fig.~\ref{fig:NaCrO2_Fig1'}(a)], are good
candidates for the physical realization of such a system. The
CrO$_2$ layers are isostructural with the CoO$_2$ layers in
superconducting Na$_x$CoO$_2$$\cdot$\textit{y}H$_2$O \cite{Takada},
and adjacent layers are well separated by intervening $A$ ions,
suggesting predominant 2D behavior in the $ab$ plane. Detailed
neutron scattering studies have only been made on LiCrO$_2$, which
orders 3D at low temperature \cite{Kadowaki}.

\begin{figure}
\includegraphics[scale=0.8,clip=true, viewport=0.0in 0.0in 6.2in
3.1in]{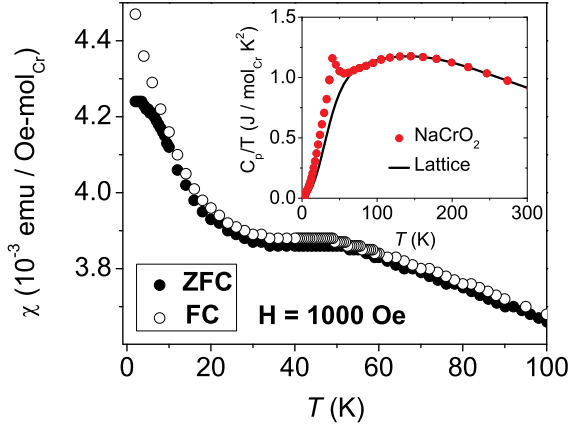} \caption{Susceptibility and specific heat.
Field cooled (FC) and zero-field cooled (ZFC) susceptibility $\chi$
measured at 1000 Oe with a SQUID magnetometer (Q.D. MPMS). Inset:
Specific heat divided by temperature $C_p$/T of NaCrO$_2$ measured
in a commercial apparatus (Q.D. PPMS) in zero magnetic field. The
magnetic contribution is obtained by subtracting out the estimated
lattice contribution (solid line) based on non-magnetic NaScO$_2$. 1 emu/(mol Oe) = 4$\pi$ 10$^{-6}$ m$^3$/mol.}
\label{fig:NaCrO2_Fig2}
\end{figure}

A 15g powder sample of NaCrO$_2$ was prepared using the method
described in \cite{Olariu}. The d.c. susceptibility $\chi$ of
NaCrO$_2$ measured up to 350K (fig.~\ref{fig:NaCrO2_Fig2}) is
consistent with earlier work up to 800K \cite{Delmas}. The
Curie-Weiss temperature $\Theta_{CW}\sim -$280K obtained from the
latter study yields an estimate of $J_{1}+J_1'=
[3k_{B}\Theta_{CW}/zS(S+1)] \sim$ $-$3 meV, where $z$ = 6 is both
the in-plane and inter-plane coordination number. The effective
magnetic moment is close to the spin-only value of 3.87$\mu_{B}$ for
spin-$\frac{3}{2}$ Cr$^{3+}$. The maximum in $\chi$ around 48 K is
broad and weakly temperature dependent, and is similarly seen in
theoretical simulations of model 2D TLAFM's \cite{Kawamura_Sus}. We
were able to rule out a spin-glass transition at this temperature
based on the absence of splitting between the field cooled (FC) and
zero-field cooled (ZFC) curves. Moreover, a.c. susceptibility
measurements showed no frequency dependence of this maximum,
implying a single relaxation time scale. The low temperature upturn
and splitting in $\chi$ is likely due to freezing of vacancy or
grain boundary spins.

The magnetic specific heat, $C_M(T)$, (inset
fig.~\ref{fig:NaCrO2_Fig2}) shows a broad maximum at 40 K, slightly
below the maximum in $\chi(T)$. The positions of these two features
are much reduced from $\Theta_{CW}(\sim -$280K), which reflects a
geometrical frustration (\textit{f}$\sim$7) induced spectral weight
downshift. Unlike the specific heat of directly stacked quasi-2D
TLAFM's of type V$X_2$ ($X$=Cl, Br) \cite{Wosnitza}, which show a
sharp lambda-type anomaly coincident with 3D magnetic ordering, the
specific heat maximum of NaCrO$_2$ is very broad \cite{Mermin}.

\begin{figure}
\includegraphics[scale=1.47,clip=true, viewport=0.0in 0.0in 2.3in
4.0in]{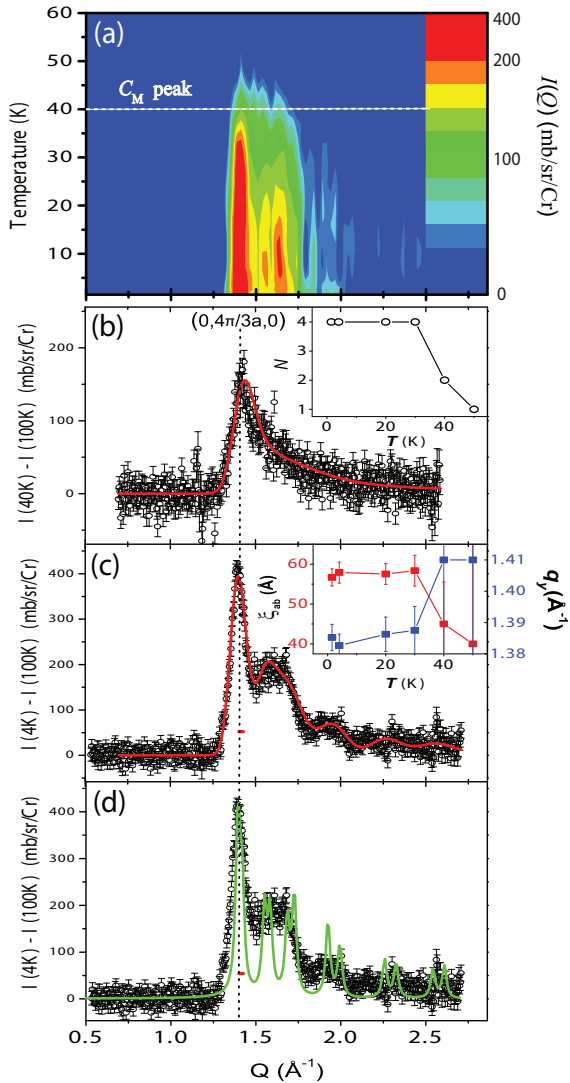} \caption{Quasi-static spin
correlations. (a) Neutron thermo-diffractogram measured using a
disk-chopper spectrometer. Intensity is displayed on a log scale to
enhance the high Q features. Dashed line indicates peak position of
magnetic specific heat $C_M$. (b) 40 K and (c) 4 K magnetic
diffraction measured in triple-axis mode using 100 K background. Red
curves are fits using methods described in the text, and the red bar
indicates the FWHM resolution width. Inset (b): Value of $N$ that
minimized $\chi^2$ at each temperature. Inset (c): Temperature
dependence of $\xi_{ab}$ and $q_y$. The green curve in (d) is a
calculated resolution limited lineshape assuming spins are 3D
long-range ordered with the fitted $\vec{q}$. Uncertainties are statistical in origin and represent one standard deviation.}
\label{fig:NaCrO2_Fig3}
\end{figure}

Magnetic neutron scattering was performed at the NIST Center for
Neutron Research. This probes the spherically averaged scattering
function $S^{\alpha\beta}(\vec{Q},\omega)$ \cite{Lovesey}

\begin{equation}\label{eq.1}
I(Q,\omega) =
r_0^2|\frac{g}{2}F(Q)|^2\int\frac{d\Omega_{\hat{Q}}}{4\pi}\sum_{\alpha\beta}(\delta_{\alpha\beta}-\hat{Q}_{\alpha}\hat{Q}_{\beta})S^{\alpha\beta}(\vec{Q},\omega)
\end{equation} where $r_0=5.38$ fm, $F(Q)$ is the magnetic form factor for
Cr$^{3+}$ and $g$ is the Land\'{e} $g$-factor. Diffraction scans
were measured on the BT2 triple-axis spectrometer with a fixed
incident and final energy of 14.7 meV. We used PG (002) reflections
for both the monochromator and the analyzer, and horizontal beam
collimations of 60$'$-40$'$-40$'$-open. The magnetic contribution,
$I(Q)$, was obtained by subtracting the 100 K data, and was
normalized to the nuclear peaks (fig.~\ref{fig:NaCrO2_Fig3}). All
magnetic peaks were resolution limited in energy ($\delta E$ = 0.07
meV), which implies that spin correlations persist on a time scale
that exceeds 0.4 ns.

The data were fit to a convolution of a Gaussian instrumental
resolution ($\delta Q=0.03$\AA$^{-1}$) with the spherical average of
magnetic scattering from a quasi-2D magnet. This is described by a
Warren function with an additional phase factor \cite{Lynn} (see
eq.~\ref{eq.2}), where $C$ is an instrumental constant,
$\vec{q}_{\parallel}$ and $\vec{q}_{\perp}$ are the components of
the ordering wave vector parallel and perpendicular to the
triangular planes respectively, and the sum is over 2D reciprocal
lattice vectors $\vec{\tau}$ [fig.~\ref{fig:NaCrO2_Fig1'}(b)]. The
ordered moment on a site $\vec{r}$ is given by $\vec{m}_{\vec{q}}
e^{i\vec{q}.\vec{r}}+\vec{m}^*_{\vec{q}} e^{-i\vec{q}.\vec{r}}$
where $\vec{m}_{\vec{q}}=(im_{qx},m_{qy},m_{qz})$. The number of
correlated layers $N$ was restricted to integer values, with
$\vec{d} = (a/(2\sqrt{3}),a/2,c/3)$ being a vector connecting
adjacent layers.

\begin{widetext}
\begin{equation}
\label{eq.2}
I(Q)=C\frac{|F(Q)|^2}{Q}\sum_{\vec{\tau}}\int^{\frac{\pi}{2}}_{-\frac{\pi}{2}}
(|\vec{m}_{\vec{q}}|^2-|\hat{Q}.\vec{m}_{\vec{q}}|^2) \times
e^{-\frac{\xi^2_{ab}}{4\pi}(Q\cos\phi-|\vec{\tau}\pm\vec{q}_{\parallel}|)^2}
\left|\sum^{N-1}_{n=0}e^{in[\vec{\tau}.\vec{d}_{\parallel}+(Q\sin\phi\mp
q_{\perp})d_{\perp}]}\right|^2d\phi
\end{equation}
\end{widetext}

Before discussing the fit results, it is worth pointing out some
differences between the powder lineshapes of directly stacked and
rhombohedrally stacked quasi-2D TLAFM's. Coherent scattering from a
single layer of 120$^{\circ}$ correlated spins comes from the
intersection of a spherical shell of radius $Q$ in reciprocal space
with Bragg rods centered at $\vec{\tau}\pm\vec{q}$, where
$\vec{q}=(0,4\pi/3a,0)$ (fig.~\ref{fig:NaCrO2_Fig1'}). For $N$=1 the
intensity along each rod is uniform, which produces the well known
saw-tooth lineshape. For $N>1$, an additional phase factor in
eq.~\ref{eq.2} generates an oscillatory intensity with period
$2\pi/d_{\perp}=6\pi/c$ along the rods. Since
$\vec{d}_{\parallel}=0$ for direct stacking, the phase factor is
independent of $\vec{\tau}$ and modulations of the saw-tooth pattern
are visible as soon as $N>1$ (see, e.g., NiGa$_2$S$_4$
\cite{Nakatsuji}). For rhombohedral stacking on the other hand, each
rod acquires an additional offset
$(\vec{\tau}.\vec{d}_{\parallel})/d_{\perp}=0$, $2\pi/c$ or $4\pi/c$
depending on the rod position [fig.~\ref{fig:NaCrO2_Fig1'}(b)]. If
intensity modulations along an individual rod are sufficiently
broad, they will be washed out when averaged over different rods.
Therefore the scattering profile can still have a pure saw-tooth
shape even for $N>1$.

Least $\chi^2$ fit results are shown in
fig.~\ref{fig:NaCrO2_Fig3}(b) and (c). For $T>40$K spin correlations
are purely 2D ($N$=1). Weak inter-plane correlations develop at 40K
($N$=2) and $N$ saturates to 4 from 30K down to 1.5K. A comparison
of the 1.5K data to the expected lineshape in the limit
$N\rightarrow\infty$ [fig.~\ref{fig:NaCrO2_Fig3}(d)] shows that
$\xi_c$ is indeed short-range. Moreover, all (00$l$) nuclear peaks
were resolution limited, which suggests that this effect is not due
to structural disorder. The in-plane correlation length $\xi_{ab}$
also increases abruptly between 50K and 30K, corresponding to 13 and
20 triangular lattice spacings respectively, and remains constant
upon further cooling. The data at all temperatures show spins
confined to the $xz$-plane (i.e. $m_{qy}\ll m_{qx},m_{qz}$), which
is consistent with electron paramagnetic resonance (EPR) results
supporting a small easy-axis anisotropy \cite{Elliston}. The average
ordered moment at 1.5K is estimated to be $\langle S^2
\rangle\approx(3/2)\int^{2.5\AA^{-1}}_{1\AA^{-1}}[I(Q)/(r_0^2|F(Q)|^2)]Q^2dQ/\int^{2.5\AA^{-1}}_{1\AA^{-1}}Q^2dQ$
=0.55(2)/Cr or $\sqrt{\langle S^2 \rangle}=0.74(1)$ which is
substantially reduced from $S$=1.5.

\begin{figure*}
\begin{center}
\includegraphics[scale=0.55,clip=true, viewport=0.0in 0in 11.0in 5.5in]{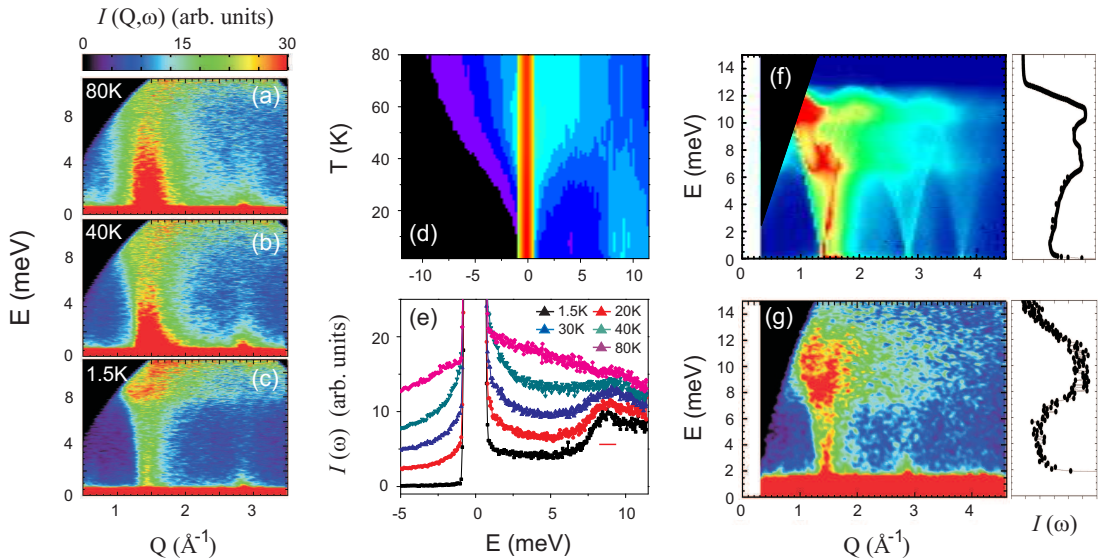}
\caption{Dynamic spin correlations. (a)-(c) Neutron scattering data
$I(Q,\omega)$ using 2.5 \AA\ incident neutrons. Panels (d) and (e)
show the temperature dependence of the magnetic density of states
$I(\omega)=\int^{2\AA^{-1}}_{1\AA^{-1}}I(Q,\omega)dQ$. Horizontal
bar shows instrument energy resolution. Panel (g) shows neutron
scattering data taken at 1.5 K measured using 1.8 \AA\ incident
neutrons, and panel (f) shows the corresponding linear spin wave
calculation.} \label{fig:NaCrO2_Fig4}
\end{center}
\end{figure*}

The development of weak $c$-axis correlations between 40K and 30K is
accompanied by a change in $\vec{q}$ [inset
fig.~\ref{fig:NaCrO2_Fig3}(c)] from (0.0(0),1.41(5),0.0(0))
$\approx$ (0,1.411,0) = (0,$4\pi/3a$,0) to (0.03(1),1.38(1),0.15(2))
respectively, and remains fairly constant down to 1.5K [inset
fig.~\ref{fig:NaCrO2_Fig3}(c)]. The ordering wave vector at 40K
points to a 120$^{\circ}$ arrangement, and is consistent with the
purely 2D nature of spin correlations and the absence of a
difference signal in the $Q\rightarrow0$ limit at this temperature.
The slight incommensuration at 30K represents a departure from
120$^{\circ}$ order which resolves the inter-layer frustration to
some extent, and naturally explains the onset of inter-layer
correlations. This supports recent $\mu$SR and NMR data
\cite{Olariu} which reveal the start of an extended crossover
regime at 41K and a maximal spin relaxation rate near 30K.

The classical ground state $\vec{q}$ of a rhombohedrally stacked
TLAFM including only $J_{1}$ and $J_1' < 3J_{1}$ was shown by
Rastelli \textit{et.al} \cite{Rastelli} to be infinitely degenerate
along helices given by
($(-2j'/\sqrt{3}a)\sin(cq_z/3)$,$4\pi/3a$+$(2j'/\sqrt{3}a)\cos(cq_z/3)$,$q_z$),
where $j'\equiv J_{1}'/J_{1}$. Quantum fluctuations select the
discrete set $q_z=2n\pi/c$ ($(2n+1)\pi/c$) for $j'$ positive
(negative), where $n$ is an integer. The $\vec{q}$ extracted from
data below 30K lies on the degenerate helix if one takes
$j'\sim-$0.1. Contrary to \cite{Olariu}, this result suggests that
inter-layer coupling is driven by exchange ($J_{1}'\sim$ 0.3 meV) as
opposed to Cr$^{3+}$ dipolar energies (3.8$\mu_B$)$^2$/$d^3 \sim$
0.06 meV. The fitted $q_z$ value 0.15(2) is close to $\pi/c=0.197$
which is consistent with $j'<0$. Therefore quantum order-by-disorder
may be responsible for the single-$\vec{q}$ low temperature magnetic
structure.

Magnetic neutron inelastic scattering experiments were performed on
the Disk Chopper Spectrometer (DCS) using 1.8 \AA\ and 2.5 \AA\
incident neutrons. Typical data sets are shown in
fig.~\ref{fig:NaCrO2_Fig4}(a)-(c). For $T>$40K, there is a
cooperative paramagnetic continuum centered at $Q$=1.4 \AA$^{-1}$
due to fluctuations of small 120$^{\circ}$-type clusters, which is a
prominent feature of triangular frustrated magnets \cite{Sato}. At
40K, there is an upward shift in the magnetic density of states
[fig.~\ref{fig:NaCrO2_Fig4}(e)], obtained by integrating
$I(Q,\omega)$ over a 1 \AA$^{-1}$ window about $Q$=1.4 \AA$^{-1}$.
Surprisingly, despite a large $\xi_{ab}$ of 45 \AA, most of the
inelastic scattering weight remains in the incoherent channel
suggesting heavily damped spin wave modes. Only below 30K, where
short-range $c$-axis correlations have been established, do
dispersive excitations appear.

In order to make contact with diffraction results, we carried out a
linear spin wave calculation of $I(Q,\omega)$ \cite{Coldea} for a
system following the Hamiltonian:

\begin{equation}
H = -J_{1}\sum_{ab_{nn}}\vec{S}_i.\vec{S}_j -
J_{1}'\sum_{c_{nn}}\vec{S}_i.\vec{S}_j
\end{equation} where $ab_{nn}$ and $c_{nn}$ denote nearest in-plane
and inter-plane neighbors respectively, $J_{1}=-$2.4 meV (AFM),
$J_1'=$0.24 meV (FM), $\vec{q}=(0.03,1.39,0.15)$, $S=\frac{3}{2}$
and spins were confined to the $xz$-plane. The result shown in
fig.~\ref{fig:NaCrO2_Fig4}(f) reproduces the main features of the
low temperature spectrum, as well as the two maxima in the density
of states. However, while there is agreement in the position of the
high energy maximum around 11 meV, the calculated position of the
low energy maximum falls short of the data by about 2 meV. This
deficit can be accounted for by including an easy-axis anisotropy
term $D$, as shown by previous numerical studies \cite{Watabe},
albeit with a $D$ value significantly larger than that deduced from
EPR measurements at 300K. One possible reason for this discrepancy
is that the single-ion $D$ increases significantly upon cooling,
which can also explain the dynamical cross-over behavior at 40K
\cite{Olariu}. Alternatively, more exotic mechanisms such as a
fluctuation induced spin wave gap can be at play (see, e.g.,
Sr$_{2}$Cu$_3$O$_{4}$Cl$_2$ \cite{Kim}). The \textit{effective}
exchange energy $J_{1}+J_1'$ obtained from spin wave analysis is
reduced by approximately 25\% compared to its \textit{bare} value
obtained from $\chi$ . This large downward renormalization is
consistent with theoretical results for the $S$=1/2 2D TLAFM
\cite{Starykh}.

To conclude, our neutron scattering measurements clearly reveal a
rare 2D antiferromagnetic phase in NaCrO$_2$ with strong spin
fluctuations over an extended temperature range. The development of
$c$-axis correlations at low temperature together with a weak
incommensurate modulation strongly suggest inter-layer
frustration as the mechanism for the stability of this unusual 2D
phase. Such a 2D magnetic phase of Mott insulators can be a host for
exotic superconductors where pairing could be mediated by the
naturally available spin fluctuations or the lifted zero-modes
unique to frustrated systems and would make an interesting variant
of the recently discovered cobaltate superconductor hydrated
Na$_x$CoO$_2$. NaCrO$_2$ has also been recently shown to be a host
for an unusual spin-driven antiferroelectric phase \cite{Seki},
which will provide important insight into how magnetic
frustration can give rise to strong coupling multiferroic behavior
\cite{Cheong}. Quite generally, our results demonstrate that
geometrical frustration provides a new avenue to stabilize 2D bulk
phases of matter.



\begin{thebibliography}{h}
\bibitem{Mermin}
      There is no violation of the Mermin-Wagner theorem since there
      is no sign of a phase transition in $C_p (T)$.
\end{thebibliography}
\end{document}